\documentclass{article}

\usepackage{times}
\usepackage[authoryear]{natbib}
\usepackage{amsmath,mathtools,amsfonts,amssymb,mathptmx}
\usepackage{longtable}
\usepackage{graphicx}
\usepackage{adjustbox}
\usepackage{subcaption}
\usepackage{tabularx}
\usepackage{float}
\usepackage[noblocks]{authblk} 
\usepackage[table]{xcolor}
\usepackage{pdfpages}
\usepackage{comment}
\usepackage{tikz}

\usepackage{docmute}
\usepackage[margin=1in]{geometry}

\usepackage{booktabs}
\usepackage{multirow}

\usepackage{hyperref}
\hypersetup{
    colorlinks=true,
    linkcolor=blue,   
    urlcolor=blue,
    citecolor=blue
    }

\usepackage[]{graphicx}
\chardef\bslash=`\\ 

\title{Implementing Response-Adaptive Randomisation in Stratified Rare-disease Trials: Design Challenges and Practical Solutions}
\author[1*]{Rajenki Das}
\author[1,2]{Nina Deliu}
\author[3,4]{Mark R Toshner}
\author[1]{Sof\'{\i}a S Villar}

\affil[1]{MRC Biostatistics Unit, University of Cambridge}
\affil[2]{MEMOTEF, Sapienza University of Rome}
\affil[3]{Victor Phillip Dahdaleh Heart \& Lung Research Institute, University of Cambridge}
\affil[4]{Royal Papworth Hospital}
\affil[*]{Corresponding author: \href{mailto:rajenki.das@mrc-bsu.cam.ac.uk}{\text{rajenki.das@mrc-bsu.cam.ac.uk}} \\ \href{mailto:rajenki.das@gmail.com}{\text{rajenki.das@gmail.com}} }
\date{ }

\begin{document}



\maketitle

\begin{abstract}
Although response-adaptive randomisation (RAR) has gained substantial attention in the literature, it still has limited use in clinical trials. Amongst other reasons, the implementation of RAR in real world trials raises important practical questions, often neglected in the technical literature. Motivated by an innovative phase-II stratified RAR rare-disease trial, this paper addresses two challenges: (1) How to ensure that RAR allocations are desirable i.e. both  acceptable and faithful to the intended probabilities, particularly in small samples? and (2) What adaptations to trigger after interim analyses in the presence of missing data? To answer (1), we propose a {\it Mapping} strategy that discretises the randomisation probabilities into a vector of allocation ratios, resulting in improved frequentist errors. Under the implementation of {\it Mapping}, we answer (2) by analysing the impact of missing data on operating characteristics in selected scenarios. Finally, we discuss additional concerns including: pooling data across trial strata, analysing the level of blinding in the trial, and reporting safety results.\end{abstract}

\maketitle                

\section{Introduction} \label{sec: Intro}
Well-designed randomised controlled trials (RCTs) have long been valued for their well-understood statistical properties and are recognised as the gold standard for conducting evidence-based clinical research to assess the efficacy of interventions. Yet, standard RCTs can demand substantial time and resources--both in terms of sample size and cost, and therefore can result impractical in cases such as rare diseases, where patient enrollment is slow and limited in size. Even in a common disease setting, many subtypes are increasingly being identified and may require personalised or stratified approaches to therapy, thus splitting the feasible number of patients that can be recruited for the overall trial into smaller groups for each subtype stratum trial~\citep{may2023rare}. Furthermore, conducting a trial with the main purpose of \textit{learning} about treatment effectiveness (as in the traditional RCTs) may be ill suited in fatal diseases, where some have suggested that the priority should be to treat trial participants as effectively as possible~\citep{may2023rare,williamson2017bayesian}. These drawbacks often prevent successful randomised experimentation and have been widely acknowledged as limiting medical innovation~\citep{bothwell16_ad}. 

Adaptive trial designs have been proposed as a means of addressing some of the practical limitations of traditional RCTs. They enable the possibility of not only enhancing the likelihood of detecting the most promising treatments without substantially increasing the sample size, but also offering expected benefit to the trial participants~\citep{bhatt16_ad}.
The fundamental characteristic of an adaptive clinical trial is to allow, according to a prespecified plan, dynamic adjustments of design features while patient enrollment is ongoing~\citep{pallmann18_ad} based on data observed at interim analysis. The first proposal of a design of this nature can be traced back to~\cite{thompson33_ad}'s idea of skewing the randomisation probabilities toward the most promising treatments according to their posterior probability of success. Due to this historical genesis, 
adaptive randomisation designs were often referred to as adaptive designs \citep{efron71_ad, lachin88_ad, rosenberger01_ad}. Although the more recent use of the term applies more generally~\citep[see e.g.,][for an overview]{bhatt16_ad,pallmann18_ad}, in this work, our focus will be on response-adaptive randomisation (RAR) designs, which prespecify how and when the randomisation probabilities should be adjusted based on the accumulated response data. We also explore how the randomisation probabilities can be used to inform early stopping rules for experimental arms.

RAR has received substantial attention in the biostatistical literature, contributing to a fertile area of methodological and theoretical research. Despite this and the recent encouragement of RAR adoption from government agencies and health authorities~\citep{fda2019_ad}, \textit{RAR uptake in clinical experimentation remains disproportionately low compared to the stream of theoretical work on this topic}~\citep{robertson2023response,baldiantognini2018_newdesign}. 
The reasons behind the gap of the RAR methodology/theory vs. the RAR in practice are diverse. 
\textit{First}, the role of RAR in clinical trials has long been and still remains a subject of active debate within biostatistics due to its potential impact on statistical inference. Bias and hypothesis testing issues, among others, have been intensively studied~\citep[see e.g.,][]{villar15_ad}, and several solutions have emerged both from the biostatistics~\citep{deliu2021efficient, bowden2017unbiased, baldiantognini2018_newdesign} and the machine learning~\citep{nie2018adaptively,deshpande2018accurate,li2022algorithms, hadad_confidence_2021} community. For a recent extensive review on the matter, we refer to~\cite{robertson2023response}, references therein, and related discussions. 
\textit{Second}, the practical debut of RAR in clinical trials, i.e., the two-armed ECMO trial~\citep{bartlett1985extracorporeal}, 
resulted in a highly controversial interpretation of its results and their generalisability due to the final extreme treatment imbalance. 
This application of RAR to a clinical trial limited the use of RAR in clinical trials for the next 20 years. \textit{Third}, the implementation phase of RAR in a real-trial context poses critical practical challenges, many of which may also apply to more traditional RCTs but which require a distinct approach when using RAR. These include, but are not limited to RAR, for example: 
\begin{itemize}
    \item[(1)] \emph{For a given vector of theoretical randomisation probabilities, how can we minimise the chances of observing undesirable treatment allocations (that is, observed allocations diverging from their theoretical counterparts beyond an acceptable level) while taking into account the impact this may have  on the design's operating characteristics?} More formally, let $\boldsymbol{\pi} = (\pi_0,\pi_1,\dots,\pi_K)$ and $\boldsymbol{\rho} = (\rho_0,\rho_1,\dots,\rho_K)$ denote the target randomisation probabilities and the observed allocation proportions, respectively, where $\pi_k$ is the randomisation probability of arm $k$ and $\rho_k$ is the proportion of participants assigned to arm $k$ (that is, $\rho_k = n_k/n$, with $n_k$ the number of participants in arm $k$ and $n$ the trial sample size). Then, our aim is to ensure $n \rho_k \approx n \pi_k$, for each arm $k$.
    Although this may be less of an issue in large-sample trials, the concern would certainly be crucial in rare-disease trials even with equal randomisation (though these issues  exacerbate with unequal sample sizes predetermined from the start or resulting from RAR). To illustrate this, consider a three-arm trial with $n=20$ and non-adaptive $\boldsymbol{\pi} = (0.5,0.4,0.1)$. From Figure \ref{fig: rho_pi} showing the distribution of $\rho$ across 10,000 simulations for each arm, it can be noted that: $\mathbb{P}(\rho_0 \leq 0.4) \approx \mathbb{P}(\rho_1 \leq 0.3) \approx 25\%$, and $\mathbb{P}(\rho_2 = 0) \geq 12\%$. This indicates a considerable likelihood of the experimental arms receiving near-equal allocation, contrary to their target probabilities, and the control arm being completely excluded. Considering the above and the fact that there will be only one trial with 20 patients, it may be both safer and statistically powerful to work on the basis of a discrete allocation ratio, such as $5:4:1$, ensuring that at least two patients are assigned to arm $k=2$.
    \begin{figure}[ht]
        \centering
        \includegraphics[scale = 0.6]{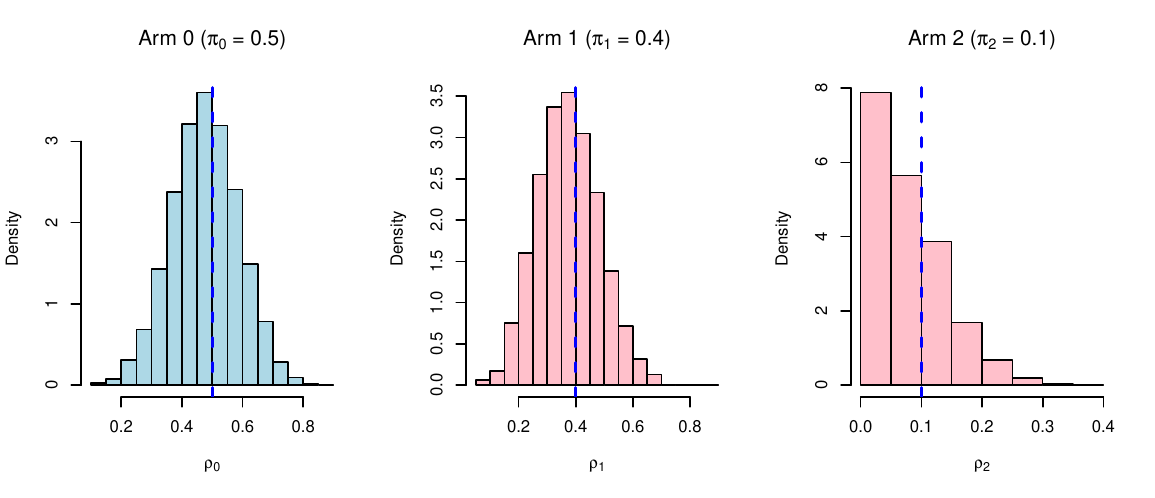}
        \caption{Empirical distribution of the observed allocation $\boldsymbol{\rho}$ of arms $k=0, 1, 2$ under the randomisation scheme $\boldsymbol{\pi} = (0.5,0.4,0.1)$. Results are averaged across 10,000 replicas of the randomisation scheme using `sample' function in R.}
        \label{fig: rho_pi}
    \end{figure}
    Furthermore, additional questions may arise when one has to practically define the allocation ratio corresponding to a target randomisation probability. This is of particular interest in RAR trials, where randomisation occurs sequentially in stages of intrinsically smaller size. For example, what should be the targeted allocation ratio corresponding to a stage-specific probability target of $\boldsymbol{\pi} = (0.5,0.4,0.1)$ with a sample size of $n = 6$? Should it be $3:2:1$ or rather $3:3:0$?

 \item[(2)] \emph{In case of missing response data that inform interim decisions in an RAR trial, when and to what extent should we allow deviations from balanced allocation in the following stage as dictated by a complete-case approach?}
 Once we address problem (1), the natural progression from there is to think about what to do if we encounter missing responses at the interim analyses. 
 A critical design decision with respect to the adaptation of the trial is whether or not to adapt the allocation towards the more promising treatment under the presence of missing responses. To the best of our knowledge, this issue has not been formally explored in the literature. 
\end{itemize}

In this work, we are specifically concerned with the research questions (1) and (2) as these are essential to the design and implementation of the motivating phase-II rare-disease RAR trial, \textit{StratosPHere 2}. 
An overview of the study is presented in Section~\ref{sec: Strato}, with the detailed protocol given in~\cite{deliu2023stratosphere}. 
For (1), we propose a {\it Mapping} rule to convert the vector of continuous target randomisation probabilities into a discrete allocation ratio object. The resulting rule preserves the randomisation properties to a chosen acceptable degree, avoids the occurrence of extreme allocation ratios by chance, and improves the operating characteristics of the original design in scenarios of interest by reducing the RAR design's variability. 
For (2), we describe a procedure for handling missing data by re-evaluating the operating characteristics and taking into account the frequency of adaptations triggered in the resulting design through simulations. Other practical challenges, going from pooled analysis to safety reporting, are also surfaced. 

Overall, with this work, we aim to discuss a set of critical practical problems along with potential solutions and recommendations, guided by our experience and collaboration with the clinical team in designing and conducting \textit{StratosPHere 2}. Our research questions are directly inspired by addressing the practical needs and the well-known difficulties of a rare disease community. We emphasise that our proposals are not meant to be regarded as universal solutions, but rather to inspire and encourage greater synergy between methodological and practical research. 
We hope our work contributes to stimulating research to increase the 
adequate adoption and implementation of adaptive designs such as RAR into clinical practice.

The remainder of this paper is structured as follows. In Section~\ref{sec: Strato}, we provide an overview of the motivating RAR trial, \textit{StratosPHere 2}, and present the preliminary notation and design setup (Section~\ref{sec: BRAR}). In Section~\ref{sec: Mapping}, we discuss the research question (1).  We explore the research question (2) in Section~\ref{sec: MissingData}. In Section~\ref{sec: FurtherChallenges}, we discuss additional challenges that are of central interest in the final analysis of our motivating trial, and, potentially, other stratified RAR trials. Final considerations and concluding remarks are given in Section~\ref{sec: discussion}.

\section{Motivating Case Study} \label{sec: Strato}

This work is motivated by practical challenges we encountered while planning the implementation of a rare-disease trial in pulmonary arterial hypertension (PAH): \textit{StratosPHere 2}~\citep{deliu2023stratosphere}. PAH is a life-threatening, progressive disorder characterised by high blood pressure in the arteries of the lungs. It affects between 15 to 50 people per million in the US and Europe and, although treatable, there is currently no cure. In this context, \textit{StratosPHere 2} represents the first-ever precision-medicine trial specifically designed to treat the causes of PAH rather than its symptoms. Importantly, it seeks to directly address devastating causes given by genetic mutations in the bone morphogenetic protein receptor type-2 (BMPR2), the most common genetic cause of familial PAH~\citep{dunmore2021approaches}.

\textit{StratosPHere 2} is a three-armed, placebo-controlled phase-II stratified RAR trial. The primary objective is to explore the efficacy of two repurposed therapies as genetic modulators of BMPR2 signaling: hydroxychloroquine and phenylbutyrate. Patients are stratified according to two specific classes of BMPR2 mutations, namely haploinsufficient (herein, Stratum A) and missense (herein, Stratum B) mutations. More formally, denoted by $T_1$ and $T_2$ the two active treatments (hydroxychloroquine and phenylbutyrate, respectively along with the standard of care) and by $C$ the control (placebo with the standard of care) treatment, trial's primary objective is to test the hypothesis that a mean increase in the primary outcome, say $\mathbb{E}(\Delta Y)$, can be achieved in the two mutation strata $s = A, B$. The primary analysis endpoint $\Delta Y$ is a measure of the engagement of the BMPR2 pathway defined as a change in the genetic expression from study entry to 8 weeks follow-up after treatment initiation. This represents a novel composite panel of validated measurements of BMPR2 target genes using quantitative PCR; we refer to Section 12 of~\cite{deliu2023stratosphere} and \textit{StratosPHere} 1 Study~\citep{jones2024stratosphere} for further details. As the two active treatments, $T_1$ and $T_2$, have distinct mechanisms of action pertinent to each stratum, the following represent the primary hypotheses to test:
\begin{align} \label{eq: hypotheses}
    \text{Stratum A}\quad\quad H_0^A\!&: \mathbb{E}\Delta Y^A_{T_1} - \mathbb{E}\Delta Y^A_C \leq 0\quad \text{vs.}\quad H_1^A\!: \mathbb{E}\Delta Y^A_{T_1} - \mathbb{E}\Delta Y^A_C > 0  \nonumber \\
    \text{Stratum B}\quad\quad H_0^B\!&: \mathbb{E}\Delta Y^B_{T_2} - \mathbb{E}\Delta Y^B_C \leq 0\quad \text{vs.}\quad H_1^B\!: \mathbb{E}\Delta Y^B_{T_2} - \mathbb{E}\Delta Y^B_C > 0. 
\end{align}

\paragraph{Primary vs. Adaptation Endpoint} The primary endpoint of the study, $\Delta Y$, is a continuous outcome, whose choice was driven by domain and statistical aspects~\citep[see][]{deliu2023stratosphere, jones2024stratosphere} and whose role is central in the final analyses. However, although study's hypotheses in Eq. \eqref{eq: hypotheses} are based on this continuous data variable, given its response-adaptive nature, an additional type of endpoint is defined with the purpose of being used to dictate the pre-planned adaptions of trial's features. This endpoint is a dichotomisation of the final continuous endpoint and it represents the \textit{adaption} endpoint. 
While the same parameter could be taken as both primary and adaptation endpoint, for the purpose of a safe and conservative adaptation in such a small trial, a binary endpoint has been considered for the latter. In fact, from the design point of view, it will result in a more conservative RAR when using a binary endpoint, and from a final analysis perspective, it would be more powerful on the continuous data. We introduce the binary indicator $\mathbf{I}(\Delta Y \geq \delta)$, where $\delta=0.3$ is the minimum meaningful change associated with a positive BMPR2 engagement. Consequently, we take as the binary {\it adaptation} endpoint the parameter $\theta_k \in [0,1]$ defined as:
\begin{align} \label{eq: adpt_end}
    \theta_k = \mathbb{E}(\Delta Y_k \geq \delta),\quad k=C, T_1, T_2.
\end{align}
This adaptation endpoint guides the randomisation probability throughout the trial, and it can be interpreted as the expected number of successes associated with each arm $k$. 

\subsection{\textit{StratosPHere 2} Design} \label{sec: BRAR}

Overall, for both strata, an expected number of $N=40$ patients is expected to be enrolled in 3 consecutive stages of randomisation, denoted by $t=1,2,3$. Specifically, we expect to recruit $n_{1,s} = 6$, $n_{2,s} = 6$ and $n_{3,s} = 8$ participants per stratum, where $n_{t,s}$ denotes the sample size of the stage $t$ block, with $t = 1, 2, 3$, and of stratum $s = A, B$. Such values are determined based on practical recruitment considerations and their frequentist properties are evaluated in simulation studies. Specifically, for the overall study duration (of at least 2 years), the rare disease allows for an expected total sample size of $n=20$ patients per stratum (and likely one patient per month) over the funding period of the trial grant. The stage-specific sizes are determined according to practical implementability and simulation evaluations aiming at maximising power. 

\begin{figure}[ht]
    \centering
    \includegraphics[clip, trim=0 6cm 0 6cm, width=1\textwidth]{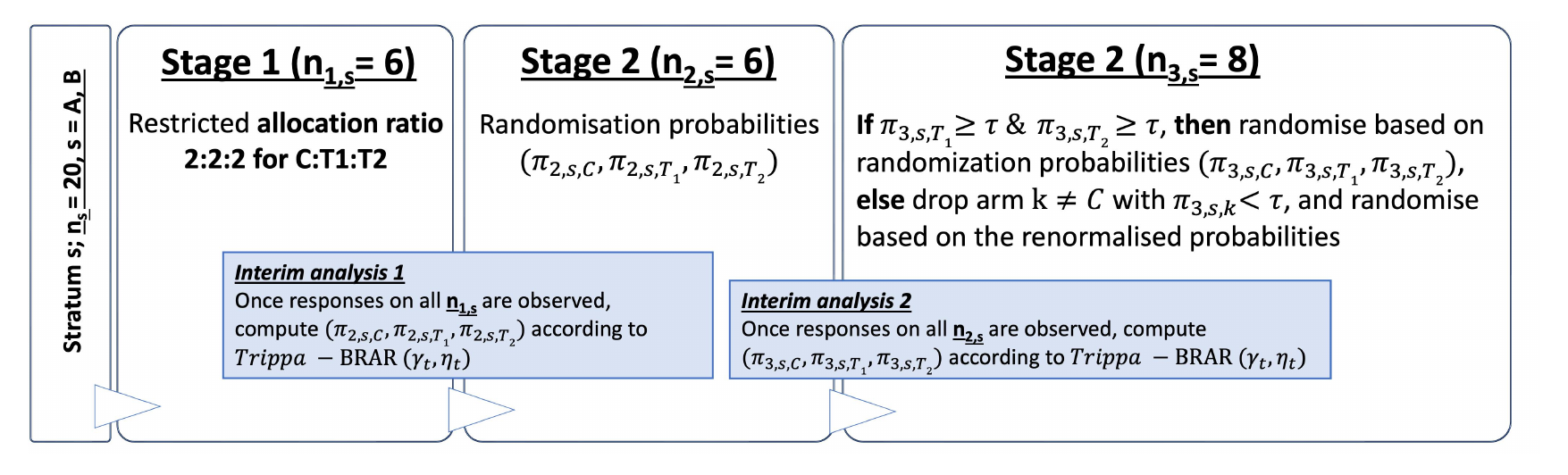}
    \caption{Schematic of the \textit{StratosPHere 2} trial design for one stratum. Threshold $\tau$ specifies the probability for arm dropping and would be disclosed at the end of the trial to preserve its integrity (e.g., avoid allocations' predictability).}
    \label{fig: StratoPH}
\end{figure}

Eligible subjects will be randomly assigned to one of the three arms ($C$, $T_1$, or $T_2$) following a Bayesian response-adaptive randomisation (BRAR) design implemented independently in each stratum $s = A, B$. In stage 1, a restricted 2:2:2 allocation ratio for $C:T_1:T_2$ is considered; this reflects a balanced randomisation probability scheme, that is, $\{\pi_{1,s,k} = 1/3; k = C, T_1, T_2\}$ with $s = A, B$. Once all responses from stage 1 are observed, a first interim analysis is performed to (possibly) update the randomisation probabilities for stage 2, say $\{\pi_{2,s,k}; k = C, T_1, T_2\}$ for all $s$. The selected design does not allow for arm dropping at this second stage. A second interim analysis (possibly) adapts the randomisation probabilities $\{\pi_{3,s,k}; k = C, T_1, T_2\}$ for the (final) stage 3 block of 8 patients in each stratum $s = A, B$. This interim accounts for stage 2 as well as stage 1 response data. In stage 3, a futile active treatment arm is allowed to be dropped if the associated randomisation probability is lower than a prespecified threshold $\tau \in [0, 0.2]$. Its precise value is documented with the trial sponsor under restricted access and will be publicly disclosed at the end of the study to minimise BRAR predictability and preserve study's integrity. 
A schematic of \textit{StratosPHere 2} design for a generic stratum $s$ is presented in Figure~\ref{fig: StratoPH}.

\paragraph{Bayesian Response-Adaptive Randomisation}
The design of \textit{StratosPHere 2} builds on the RAR allocation rule proposed in~\cite{trippa2012bayesian} and further discussed in \citep{wason2014comparison}. It adopts a Bayesian framework for adjusting the randomisation probabilities at each interim analysis based on the accumulated response data up to that point. 
Let $\mathcal{D}_{n_{\overline{t-1}, s}}$ be all the observed data (assigned arms and corresponding outcomes or responses) from the $n_{\overline{t-1},s}$ participants of stratum $s$, where $\overline{t-1}$ refers to all accumulated stages $\{1,\dots,t-1\}$. For example, at stage $t=3$, $\overline{t-1} = \{1,2\}$ and includes $n_{1,s} + n_{2,s}$ patients. Then, Trippa's rule defines the stage $t$ randomisation probabilities for stratum $s$, say $\{\pi^{\text{Trippa}}_{t, s, k}; k=C, T_1, T_2\}$, as:
\begin{align} \label{eq: Trippa}
    \pi^{\text{Trippa}}_{t, s, k} (\gamma_t, \eta_t) \propto 
    \begin{dcases}
        \frac{p(\theta_k > \theta_C \mid \mathcal{D}_{n_{\overline{t-1}, s}})^{\gamma_t}}{\sum_{j\in \{ T_1, T_2\}} p(\theta_j > \theta_C \mid \mathcal{D}_{n_{\overline{t-1}, s}})^{\gamma_t}}\quad &\text{if}\ k=T_1,T_2,\\
        \frac{1}{K}\exp{(\max(n_{\overline{t-1},s, T_1}, n_{\overline{t-1},s, T_2}) - n_{\overline{t-1},s, C})^{\eta_t}}\quad &\text{if}\ k=C,
    \end{dcases}
\end{align}
where $n_{t,s,k}$ represents the number of individuals assigned to arm $k$ in stratum $s$ up to time $t$, $K = 3$ is the number of study arms, and $\theta_k$ reflects the adaptation endpoint associated with arm $k$, as defined in Eq.~\eqref{eq: adpt_end}. The two stage-varying hyper-parameters $\gamma_t$ and $\eta_t$ are introduced to modulate, respectively: (i) the current and final allocation imbalance between the two active treatment arms $T_1$ and $T_2$ and; (ii) the overall allocation of the control arm $C$ at the end of the study. We will refer to this rule as \textit{Trippa-BRAR}$(\gamma_t, \eta_t)$ and to the associated design as \textit{Control Protected}. In \textit{StratosPHere 2}, $\eta_t$ is tuned so as to guarantee a minimum allocation of the control arm of around $1/K \approx 0.33$. 

Note that \textit{Trippa-BRAR}$(\gamma_t, \eta_t)$ can be viewed as an extension of the popular Thompson sampling~\citep[TS;][]{thompson33_ad} rule, recognised as the first prototype of a BRAR design. Here, randomisation probabilities are expressed in terms of their posterior probability of being associated with the maximum expected outcome $\theta_k$, that is,
\begin{align} \label{eq: TS}
    \pi^{{TS}}_{t, s, k} (\gamma) \propto 
    \frac{p(\theta_k > \theta_k', k \neq k' \mid \mathcal{D}_{n_{\overline{t-1}, s}})^{\gamma}}{\sum_{j\in \{ C, T_1, T_2\}} p(\theta_j > \theta_j', j \neq j' \mid \mathcal{D}_{n_{\overline{t-1}, s}})^{\gamma}}\quad k=C,T_1,T_2,
\end{align}
where $\gamma$ is a positive tuning hyper-parameter introduced by~\cite{thall07_ad} for stabilising the randomisation probabilities.When $\gamma = 1$, Eq.~\eqref{eq: TS} reduces to vanilla TS, as first proposed in~\cite{thompson33_ad}. We term this rule \textit{TS-BRAR$(\gamma)$} and the associated design \textit{Unrestricted}, given that no restrictions are placed on the randomisation probabilities, which are uniquely guided by the observed responses. For example, compared to \textit{Trippa-BRAR}$(\gamma_t, \eta_t)$, \textit{TS-BRAR$(\gamma)$} does not impose any restrictions on the control arm. Finally, note that taking $\gamma = 0$, gives us conventional balanced randomisation. 

Being a Bayesian framework, we make use of prior distributions on the unknown parameters $\theta_k$ for $k = C, T_1, T_2$. To reflect the novelty of the study, we assume Beta($1,1$) prior distributions for all arms. 

\subsection{Operating Characteristics} We evaluated by extensive simulations the frequentist properties (type-I error and power) as well as the probability of patients receiving a superior arm when this exists. As detailed in~\cite{deliu2023stratosphere}, all evaluations for final analysis, were implemented using a fully non-parametric approach based on a bootstrap resampling technique applied on response data -- corresponding to cases (patients) but with standard of care treatment, available from a preliminary phase~\citep[\textit{StratosPHere 1};][]{jones2024stratosphere} and using a one-sided Wilcoxon test. The non-parametric approach is motivated by the small sample and the non-normally distributed data. As illustrated in Table~\ref{tab:mapping_OCs}, satisfactory results are expected with the proposed sample size. Simulation studies resulted in almost $80\%$ power under a $10\%$ type-I error control in each stratum of $n=20$ patients for the BRAR design. Importantly, it highlights the benefits of a BRAR design over a fixed strategy in both allocating the most promising arms (column 5), and achieving an increased power in multi-armed cases (column 3). 

\section{Mapping of Allocation Probabilities to Allocation Ratios}
\label{sec: Mapping}

Randomisation is an essential component of a trial design. In complex trial settings, careful considerations are required from the implementation side to choose an appropriate randomisation procedure (be this adaptive or not). For example, in small trials, particular emphasis is placed on avoiding {\it undesirable} allocations, which can occur when targeting a particular allocation ratio. 
Furthermore, with the spread of adaptive designs, there is an increasing need for randomisation methods that allow unequal allocations. The rationale for using unequal allocation is provided by \citet{peckham2015use} and \citet{dumville2006use}, underlining factors such as cost considerations or ethical concerns. 
However, methodological challenges of randomisation in small samples remain; see, e.g.,~\cite{berger2021roadmap} and~\cite{van2019merged}, who discuss the permuted-block design and put forward merged-block designs to lower predictability. To achieve a target allocation ratio,~\cite{sverdlov2019implementing} propose a truncated multinomial design where participants are randomised to treatments according to a multinomial distribution until the target allocation number is reached for each treatment. \cite{kuznetsova2012preserving} also point out the importance for robust randomisation systems to preserve the targeted allocation ratio, discussing and calling for alternative allocation procedures, such as biased coin randomisation, that better approximate the allocation ratio in small samples, while reducing the selection bias in open-label studies. Designs such as the big-stick design \citep{soares1983some}, the block-urn design \citep{zhao2011block}, the maximal procedure \citep{berger2003minimizing}, or the brick-tunnel design \citep{kuznetsova2011brick}, among others, are some of the randomisation methods that effectively achieve unequal treatment allocations, \textit{once the allocation ratio has already been determined}. In parallel, one also faces the problem of determining these practical allocation ratios, by which we mean the allocation ratios are desirable given the allocation probabilities and they adhere to the constraints of the trial design. Existing proposals, e.g., \cite{tymofyeyev2007implementing}, discuss how allocations can be optimised in order to maximise power; however, resulting values are typically defined on a continuous range and their translation in practice is underdeveloped. Therefore, there is gap between methods to define ideal allocation ratios reflecting the underlying design's goal and the existing procedures to best implementing resulting ratios into a feasible the randomisation system. This is particularly true for BRAR settings, which is the most used RAR in practice so far \citep{pin2025github}. 

To summarise, managing unequal allocation ratios with small sample sizes is a challenge within itself (even outside of RAR designs). Particularly, BRAR designs pose additional challenges since instead of working with whole number ratio allocation weights, it primarily returns (continuous) allocation probabilities. This makes it more challenging to achieve the desired allocation, as described within the research problem 1 in the introduction. 

In this paper, we aim to fill the aforementioned gap. Specifically, we propose a method to map the (continuous) allocation probabilities derived according to the design's allocation rule, e.g., Thompson sampling or Trippa's rule, to a suitable ratio using probability thresholds which act as boundaries for decision making for the adaptive design. We call this method {\it Mapping}. As a result, the randomisation process preserves its adaptability and operating characteristics up to the level of avoiding pre-established undesirable allocations. 
Furthermore, {\it Mapping} simultaneously covers both the both the determination of the target ratio and its implementation. This guarantees that the randomised sequence (randomised list) adheres to the constraints of the design and maintains the expected allocation ratios.  

This section focuses on research problem (1) as outlined in the Introduction. First, we provide a comparison amongst selected designs considered for \textit{StratosPHere 2} differing in their degree of constraints on the allocation probabilities. We will refer to them as the \textit{Unmapped} RAR design, since their continuous randomisation probabilities are directly used to determine the allocations. Then, we consider the baseline design to be \textit{StratosPHere 2} (as illustrated in Figure~\ref{fig: StratoPH} and discussed in the study protocol; \cite{deliu2023stratosphere}). Finally, we present {\it Mapped} designs which includes Permuted Block Design and our proposed designs that translate the continuous allocation probabilities to discrete target allocations. 
A preliminary summary of the comparators is given in Table~\ref{tab:RAR_designs}. 

Since each design is implemented independently within each stratum, we outline our proposal with reference to a single stratum. The same approach can be replicated for the other stratum. 
We recall that each stratum is structured in three stages $t=1,2,3$ with sizes $n_t$ determined according to practical recruitment considerations: $n_{1} = 6$, $n_{2} = 6$ and $n_{3} = 8$. At the end of each stage $t-1$, i.e., during the interim analysis, the BRAR updates the vector of randomisation probabilities for the allocation of the treatments for the next stage: $\{\pi_{t,s,k}; k = C, T_1, T_2\}$.

\subsection{Unmapped Designs}
We consider three types of \textit{Unmapped} designs -- first one is \textit{Fixed Equal Randomisation} where there is a $1/K$ probability ($K$ is the total number of arms) of getting a treatment arm getting assigned to a patient. 
The other two designs include:\textit{Unrestricted} where we do not impose any of the trial constraints and the allocations are given by Thompson sampling probabilities \textit{TS-BRAR$(\gamma)$}  as in Eq.~\eqref{eq: TS} where $\gamma = 1$, and \textit{Control Protected} where the allocation rule is \textit{Trippa-BRAR}$(\gamma_t, \eta_t)$ as given in Eq.~\eqref{eq: Trippa} -- this ensures a minimum allocation of the control arm.  
Compared to \textit{Control Protected} design, \textit{StratosPHere 2} design based on the motivating study has the additional trial restrictions per stage, namely, a restricted randomisation with an allocation ratio of $2:2:2$ at stage 1 and no arm dropping allowed at stage 2, while early arm dropping is allowed in stage 3 (see Figure~\ref{fig: StratoPH}). Given its relevance, we shall use it as a baseline design for constructing our {\it Mapping} strategy. An outline of the compared designs is reported in Table~\ref{tab:RAR_designs}.

These designs include direct implementation according to the continuous randomisation probabilities estimated at each interim analysis (end of stages 1 and 2; see Figure~\ref{fig: StratoPH}) with no guaranteed properties of the resulting allocation sequences. As an example, think of assigning the second stage of 6 patients to the three treatment arms by rolling a fair 3-sided die i.e. with probability of 1/3 each. Since the allocation is determined randomly only once, there is no guarantee that the assignments will result in a 2:2:2 allocation across the treatments due to the stochastic nature of the process; thus, the resulting sequence (randomisation list) of treatments may deviate from the expected or desired allocation.

\begin{table}[h]
    \centering
    \adjustbox{max width = \linewidth}{
    \begin{tabular}{l|l|c|c|c|c}
    \toprule
          Design type & Design name & Allocation rule & Stage 1 & Stage 2 & Stage 3\\
    \midrule
         \multirow{3}{*}{\textit{Unmapped}} & \textit{Fixed Equal Randomisation} & -- & -- & -- & -- \\
         & \textit{Unrestricted} & \textit{TS-BRAR$(\gamma)$}                            & -- & -- & -- \\
         & \textit{Control Protected} & \textit{Trippa-BRAR}$(\gamma_t, \eta_t)$ & -- & -- & --  \\
        
                                   \hline
Baseline & \textit{StratosPHere 2} & \textit{Trippa-BRAR}$(\gamma_t, \eta_t)$  & 2 : 2 : 2  & No arm dropping & -- \\
\hline
                                
         \multirow{3}{*}{\textit{Mapped}} 
         & \textit{Permuted Block} & -- & 2 : 2 : 2 & 2 : 2 : 2 & 2 : 3 : 3 \\
         & \textit{Mapped-$\alpha$} & \textit{Trippa-BRAR}$(\gamma_t, \eta_t)$ & 2 : 2 : 2  & No arm dropping;  $\#C = 2$ &  $\#C = 2$ 
         \\
         & \textit{Mapped-$\beta$} & \textit{Trippa-BRAR}$(\gamma_t, \eta_t)$ & 2 : 2 : 2 & No arm dropping; $\#C = 2$  & $\#C = 2$\\
    \bottomrule
    \end{tabular}
    }
    \caption{Taxonomy of the evaluated BRAR designs, with corresponding allocation rule and restrictions per stage. $\#C = 2$ indicates the exact number of controls allocated.  
    }
    \label{tab:RAR_designs}
\end{table}

\paragraph{Implementation of \textit{Unmapped} vs. \textit{Mapped} Designs}
Unmapped procedures require the randomisation system to use probabilistic assignment methodology i.e., utilise probabilities derived from the BRAR algorithm along with some type of random number generator and logic to perform patient assignments. Implementing this within a randomisation system is complex and requires advanced software/coding by randomisation system providers. This level of complexity can limit the number of providers capable of supporting such systems. Whereas for \textit{Mapped} designs,
the derived discrete allocations can be implemented within the standard randomisation schedules (e.g., randomisation list with permuted blocks containing the included treatments and allocations). These standard randomisation schedules can be used in any randomisation system.

\subsection{Mapped Designs}
Now, we describe our proposal for the \textit{Mapped} versions of the \textit{Unmapped} designs. In practice, {\it Mapping} can be viewed as an intermediate step to link the adaptive design's definition to its implementation in a concrete trial setting. As such, {\it Mapping} involves a decision rule to translate the continuous randomisation probabilities derived in the \textit{Unmapped} designs at the interim analyses into a target vector of discrete allocation ratios of the form $R_0:R_1: \dots :R_K$ for a trial with $K$ arms. Its fundamental principle is to define a set of probability thresholds to split the continuous probability space into discrete categories that match possible values of the allocation ratios $R_0:R_1:\dots :R_K$. This ensures an efficient allocation of treatments to the small population where we want to avoid the possibility of undesired allocations, while maintaining interpretability of the process. 

In Table \ref{tab:RAR_designs}, we have listed three designs as \textit{Mapped} -- first of which is the \textit{Permuted Block Design} where the allocation ratio is fixed to maintain balance in the three different stages. The other two designs, which are discussed in the rest of the section, allow more room to deviate from the balanced allocations if needed.

\paragraph{Definition of the Discrete Allocation Ratio Space} The allocation ratio is directly driven by the allocation probabilities dictated by the underlying \textit{Unmapped} design. Given its practical relevance, we will focus on the baseline design \textit{StratosPHere 2} where we are interested in the $R_0:R_1:R_2$ possible allocation ratios. An important step 
is establishing discrete ``adaptation" categories within the {\it Mapping} strategy or 
categories which represent 
how much a promising arm could be given preference given observed data and by design. 

Specifically, in our implementation, we consider the following five decision categories depending on the stage of the trial and the data observed: \textit{Drop}, \textit{Disfavour}, \textit{Balance}, \textit{Favour}, and \textit{Keep}. The first two categories refer to a situation in which an active treatment arm shows unpromising results relative to the other active arm, as opposed to the last two categories; category \textit{Balance} indicates a case of relative indifference between the two active treatment arms. Note that, for a given arm, these categories are mutually exclusive -- if an arm is determined as \textit{Drop} and it is the only arm in this category, it will be excluded from further allocation and the \textit{Disfavour} allocation ratios will not apply. To resemble the design of \textit{StratosPHere 2} (which has a protection on the control treatment), we start by fixing the number of allocations to control $R_0$ at 2 for each stage in the \textit{Mapped} versions. This ensures that the pre-defined final allocation of approximately $1/K \approx 0.3 = 6/20$ for the control arm is preserved. 
Therefore, we now focus exclusively on categorising the mapping for the active treatment arm(s). 

Guided by the \textit{StratosPHere 2} design, in Table~\ref{tab:mapping_ratioslist}, we present possible decisions that can be made regarding the allocation ratios for $C : T_1 : T_2$, for each stage and for each category. At stage 1, in the absence of sufficient information about the treatment effects, we start with an equal allocation ratio (or, equivalently, we set the \textit{Balance} category for all arms). In stage 2, we allow for mild skewing of the allocation ratio towards the more promising arm, but \textit{Drop} or \textit{Keep} is not an option here. \textit{Drop} i.e., removing an active arm, or \textit{Keep} i.e., dropping the other active arm is an additional possible choice given the data only in stage 3. 

\begin{table}[ht]
    \centering
    \begin{tabularx}{\textwidth}{ llXXX}
    \toprule
        \textbf{Category} & \textbf{Description} & \textbf{Stage 1}   & \textbf{Stage 2}   & \textbf{Stage 3} \\
      \hline
         \textit{Drop}    &  $T_1$ can be dropped     & Never        & Never        & 2 : 0 : 6 \\  
                        \midrule
         \textit{Disfavour}  &  $T_1$ can be disfavoured, but not dropped & Never        & 2 : 1 : 3 & 2 : 1 : 5 \\
                        &           &           & & 2 : 2 : 4 \\
                        \midrule
         \textit{Balance}     &   The active arms can be allocated equally  & 2 : 2 : 2 & 2 : 2 : 2 & 2 : 3 : 3 \\
         \midrule
         \textit{Favour}    &   $T_1$ can be favoured, without dropping the other & Never        & 2 : 3 : 1 & 2 : 4 : 2 \\
                        &           &           & & 2 : 5 : 1 \\
             \midrule
         \textit{Keep}     &  $T_1$ can be kept, while dropping the other    & Never        & Never        & 2 : 6 : 0 \\
      \bottomrule
\end{tabularx}
    \caption{Possible options for the allocation ratios for $C : T_1 : T_2$ at each stage of the trial based on the categories of the active treatment arms. ``Never'' refers to a situation in which that category is not an option (e.g., at stage 1, we \textit{always} choose \textit{Balance}, with arms allocated based on a $2:2:2$ ratio, and ``Never'' the other categories).}
    \label{tab:mapping_ratioslist}
\end{table}


Given the categories per interim stage, one can define multiple types of mapped designs depending on the number of thresholds that are considered per stage on the continuous probability space. In this paper, we discuss two types of \textit{Mapped} designs which we will call \textit{Mapped-$\alpha$} and \textit{Mapped-$\beta$}. In \textit{Mapped-$\alpha$}, we introduce a probability threshold in stage 2 to distinguish between \textit{Disfavour} and \textit{Favour}, followed by two distinct probability thresholds in stage 3 to distinguish between two of the consecutive categories of \textit{Drop}, \textit{Disfavour}, \textit{Favour} and \textit{Keep}. Let $\mathcal{M}^{m}_{t}$ denote the mapping domains for \textit{Mapped} design $m$ and stage $t$ of the trial. Thus, $\mathcal{M}^{\alpha}_2 = \{ \textit{Disfavour}, \textit{Favour}\}$ and $\mathcal{M}^{\alpha}_3 = \{ \textit{Drop}, \textit{Disfavour}, \textit{Favour}, \textit{Keep}\}$ at stages 2 and 3 respectively, while $\mathcal{M}^{\alpha}_1 = \{ \textit{Balance} \}$. Notice that this design does not allow for any balance region on the decision line for the categorisation of an active treatment arm. This motivates the consideration of the second \textit{Mapped} design, a refined one called \textit{Mapped-$\beta$}, which accommodates for a \textit{Balance} option (i.e. no adaptation) on the decision line. That is, \textit{Mapped-$\beta$} has the following mapping domains while preserving $\mathcal{M}^{\beta}_1 = \mathcal{M}^{\alpha}_1 = \{ \textit{Balance} \}$: $\mathcal{M}^{\beta}_2 = \{ \textit{Disfavour}, \textit{Balance}, \textit{Favour}\}$ and $\mathcal{M}^{\beta}_3 = \{ \textit{Drop}, \textit{Disfavour}, \textit{Balance}, \textit{Favour}, \textit{Keep}\}$. This is achieved by taking a higher number of probability thresholds as shown in Figure~\ref{fig: mapping_stages} which provides an illustrative schematic of the proposed \textit{Mapped} designs. We can notice that once the \textit{Balance} region converges to an empty set (as depicted by the arrows), then \textit{Mapped-$\beta$} converges to \textit{Mapped-$\alpha$}. An important consideration is if it is important to keep the \textit{Balance} option available throughout the trial, in which case one may want to favour the design with higher granularity. Intuitively, the higher granularity mapping requires a higher level of ``evidence'' in the observed probabilities in order to deviate from a balanced allocation. It is also a pragmatic design in the sense that it will allow for not changing the design when data is not favourable enough to do so. 

Thus, to summarise, \textit{Mapped} designs implement the usage of thresholds $p_{t,i}$ at stage $t$ of the trial indexed at $i$ for discretising the decision line. \textit{Mapped-$\beta$} has two thresholds at stage 2: $p_{2,1}'$ and $p_{2,1}''$ and four thresholds at stage 3 $p_{3,1},p_{3,2}', p_{3,2}'', p_{3,3}$. When the hyphenated thresholds are equal i.e., $p_{2,1}'$ = $p_{2,1}''$ and $p_{3,2}'= p_{3,2}''$ then the design simplifies to \textit{Mapped-$\alpha$} eliminating the {\it Balance} region on the decision line.

\begin{figure}[h]
    \begin{center}
    \fbox{\textit{Mapped-$\beta$}} 
    \vspace{0.18cm}

    \begin{minipage}{0.375\textwidth} 
        \centering
        \begin{tikzpicture}[scale=0.6] 
        \scriptsize
            \draw[->] (0,0) -- (10,0) node[right] {};
            \node[below] at (0, 0) {0};
            \node[below] at (10, 0) {1};

            \node[below] at (3, -0.1) {$p'_{2,1}$};
            \draw[dashed, thick] (3, 0) -- (3, 1.5);
            \node[above] at (1.5, 1.5) {Disfavour};
            \draw[->, dashed] (1.5, 1.2) -- (2.9, 1.2);

            \node[below] at (7, -0.1) {$p''_{2,1}$};
            \draw[dashed, thick] (7, 0) -- (7, 1.5);
            \node[above] at (5, 1.5) {Balance};

            \node[above] at (9, 1.5) {Favour};
            \draw[<-, dashed] (7.1, 1.2) -- (9, 1.2);
        \end{tikzpicture}
        \\ 
        Stage 2
    \end{minipage}%
    \hspace{0.16\textwidth} 
    \begin{minipage}{0.375\textwidth}
        \centering
        \begin{tikzpicture}[scale=0.6] 
        \scriptsize
            \draw[->] (0,0) -- (10,0) node[right] {};
            \node[below] at (0, 0) {0};
            \node[below] at (10, 0) {1};

            \node[below] at (1.8, -0.1) {$p_{3,1}$};
            \draw[dashed, thick] (1.8, 0) -- (1.8, 1.5);
            \node[above] at (1, 1.45) {Drop};

            \node[below] at (3.9, -0.1) {$p'_{3,2}$};
            \draw[dashed, thick] (3.9, 0) -- (3.9, 1.5);
            \node[above] at (2.8, 1.5) {Disfavour};
            \draw[->, dashed] (2, 1.2) -- (3.8, 1.2);

            \node[below] at (6.1, -0.1) {$p''_{3,2}$};
            \draw[dashed, thick] (6.1, 0) -- (6.1, 1.5);
            \node[above] at (5, 1.5) {Balance};

            \node[below] at (7.6, -0.1) {$p_{3,3}$};
            \draw[dashed, thick] (7.6, 0) -- (7.6, 1.5);
            \node[above] at (6.8, 1.5) {Favour};
            \draw[<-, dashed] (6.2, 1.2) -- (7.5, 1.2);

            \node[above] at (9, 1.45) {Keep};
        \end{tikzpicture}
        \\ 
        Stage 3
    \end{minipage}

    \vspace{0.5cm} 

    \fbox{\textit{Mapped-$\alpha$}} 
    \vspace{0.18cm}

    \begin{minipage}{0.375\textwidth}
        \centering
        \begin{tikzpicture}[scale=0.6] 
        \scriptsize
            \draw[->] (0,0) -- (10,0) node[right] {};
            \node[below] at (0, 0) {0};
            \node[below] at (10, 0) {1};

            \node[below] at (4, -0.1) {$p_{2,1}$};
            \draw[dashed, thick] (4, 0) -- (4, 1.5);
            \node[above] at (2.5, 1.5) {Disfavour};

            \node[above] at (5.5, 1.5) {Favour};
        \end{tikzpicture}
        \\ 
        Stage 2
    \end{minipage}%
    \hspace{0.16\textwidth}
    \begin{minipage}{0.375\textwidth}
        \centering
        \begin{tikzpicture}[scale=0.6] 
        \scriptsize
            \draw[->] (0,0) -- (10,0) node[right] {};
            \node[below] at (0, 0) {0};
            \node[below] at (10, 0) {1};

            \node[below] at (1.8, -0.1) {$p_{3,1}$};
            \draw[dashed, thick] (1.8, 0) -- (1.8, 1.5);
            \node[above] at (1, 1.45) {Drop};

            \node[below] at (4.1, -0.1) {$p_{3,2}$};
            \draw[dashed, thick] (4.1, 0) -- (4.1, 1.5);
            \node[above] at (2.8, 1.5) {Disfavour};

            \node[below] at (6.1, -0.1) {$p_{3,3}$};
            \draw[dashed, thick] (6.1, 0) -- (6.1, 1.5);
            \node[above] at (5.1, 1.5) {Favour};

            \node[above] at (7.2, 1.5) {Keep};
        \end{tikzpicture}
        \\ 
        Stage 3
    \end{minipage}
\end{center}
\caption{Schematic of the proposed \textit{Mapping} designs. Closing the {\it Balance} region in the top-row design \textit{Mapped-$\beta$} converts it into the \textit{Mapped-$\alpha$} design as shown in the bottom row.
  }
  \label{fig: mapping_stages}
\end{figure}

Once a category for an active arm has been chosen based on the thresholds by the \textit{Mapped} design (threshold selection is discussed later), the corresponding discrete allocation ratio is selected from the options listed in Table~\ref{tab:mapping_ratioslist}. If two active arms fall in the same category, say \textit{Disfavour}, then we opt for the balanced allocation of that stage. If the category has two possible allocation ratios, then either  of them is chosen at random with equal probability. 

\paragraph{Definition of the Mapping functions}

We now give a formal description of our proposed approach, encompassing both the decision rule and the allocation strategy. At the end of stage 1 and 2, we observe some interim data and obtain a vector of allocation probabilities $\boldsymbol{\pi} = [\pi_{C}, \pi_{T_1}, \pi_{T_2}]$ where $C$, $T_1$, and $T_2$ denote Control, Treatment 1 and Treatment 2 respectively. For a {\it Mapping} $m$ in stage $t$, for each active treatment arm $k = T_1, T_2$, we define a decision rule $Dec^{m}_t(\pi_k)$
as a function of the allocation probabilities.
This determines the allocation decision of assigning the active treatment arm to an adaptation, and based on the decision, the final allocation ratio gets specified. 

For example, the representation of Figure~\ref{fig: mapping_stages} can be formalised through a mapping function $Dec^{m}_t(\pi_k)$ when \textit{Mapped} design $m$ is $\beta$ as:
$$
Dec^{\beta}_{t}(\pi_k): [0,1] \to \mathcal{M}^{\beta}_t,
$$
where $[0,1]$ is the domain of $\pi_k$ while $\mathcal{M}^{\beta}_t$ is the stage $t$ domain of the discrete allocation ratio introduced by the mapping. Specifically, $\mathcal{M}^{\beta}_2 = \{Disfavour, Balance, Favour\}$ and $\mathcal{M}^{\beta}_3 = \{Drop, Disfavour, Balance, Favour, Keep\}$. 
More generally, for a set $\mathcal{M}_t^m = \{{Category}_1, \dots, {Category}_J\}$ of adaptation categories under a mapping $m$ at stage $t$, the allocation probability space $[0,1]$ is partitioned into subintervals by thresholds $\{p_{t,0}, p_{t,1}, \dots, p_{t,J}\}$ such that $0 = p_{t,0} < p_{t,1} < \dots < p_{t,J} = 1$, and the decision rule outputs one of the given categories $j$:\\
$${Dec}^m_t(\pi_k) = {Category}_j \quad \text{if} \quad  \pi_k \in [p_{t,j-1}, p_{t,j}).$$

Once the mapping rule outputs an adaptation category for each active treatment arm, we can finalise the allocation ratio based on the number of arms in each category. This is defined using an auxiliary function \( Alloc^m_t(\lvert T^{Category} \rvert) \) where $\lvert X \rvert$ denotes the cardinality of the active arms assigned to a given adaptation category and $t \in \{2,3\}$ represents the trial stage. The bar notation above a category name, e.g., $T^{\overline{Balance}}$, indicates the complement set for that active arm -- meaning, not in that category. Formally, for the more granular \textit{Mapped} design $m = \beta$, we define it as:
\begin{align*}
Alloc^{\beta}_{2}(\lvert T^{Category} \rvert) =
    &\begin{cases}
        2 : 1 : 3 :: C : T^{Disfavour} : T^{\overline{Disfavour}} & \text{if } \lvert T^{Disfavour} \rvert = 1 \\
        2 : 2 : 2 :: C : T^{Balance} : T^{\overline{{Balance}}} & \text{if } \lvert T^{Balance} \rvert = 2 \\
        2 : 3 : 1 :: C : T^{Favour} : T^{\overline{Favour}} & \text{if } \lvert T^{Favour} \rvert = 1 \\
        2 : 2 : 2 :: C : T_1 : T_2 & \text{otherwise}
    \end{cases}; \\
Alloc^{\beta}_{3}(\lvert T^{Category} \rvert) = 
    & 
    \begin{cases}
        2 : 0 : 6 :: C : T^{Drop} :T^{\overline{Drop}} & \text{if } \lvert T^{Drop} \rvert = 1 \\
        2 : 1 : 5 \text{ or } 2 : 2 : 4 :: C : T^{Disfavour} : T^{\overline{Disfavour}} & \text{if } \lvert T^{Disfavour} \rvert = 1 \\
        2 : 5 : 1 \text{ or } 2 : 4 : 2 :: C : T^{Favour} : T^{\overline{Favour}} & \text{if } \lvert T^{Favour} \rvert = 1 \\
        2 : 3 : 3 :: C : T_1 : T_2 & \text{if } \lvert T^{Balance} \rvert = 1 \\
        2 : 6 : 0 :: C : T^{Keep} : T^{\overline{Keep}} & \text{if } \lvert T^{Keep} \rvert = 1 \\
        2 : 3 : 3 :: C : T_1 : T_2 & \text{otherwise}
\end{cases}.
\end{align*}

\paragraph{Determination of the \emph{Mapping} Thresholds} The implementation of {\it Mapping} requires fixing the set of probability thresholds $\{p_{t,0}, p_{t,1}, \dots, p_{t,J}\}$. Here we do this empirically by assessing the impact of possible threshold values on resulting operating characteristics of the adopted mapped design. 
For simplicity we illustrate this process using the \textit{Mapped-$\alpha$} design but a similar (yet more complex) process can be used for the \textit{Mapped-$\beta$} design. 

Thresholds are selected per stage using simulations, with calibration of an additional parameter that captures the level of \textit{adaptability} achieved across simulations at each stage. With \textit{adaptability} we refer to 
decisions for ``Favouring'' or ``Dropping'' an active arm, thereby deviating from a balanced allocation between the two active treatment arms. 
In our motivating trial, this occurs when the allocation ratios in stage 2 and stage 3 deviate from the initial 2:2:2 and 2:3:3 due to an active arm being favoured or dropped. Our goal is to determine thresholds such that adaptations are less often triggered under the null hypothesis (no difference in treatment arms) while also resulting in good operating characteristics, particularly in terms of power (under the alternative hypothesis: superiority of an active arm).

\begin{figure}[h]
     \begin{subfigure}[b]{0.5\textwidth}
         \centering
         \includegraphics[width=\linewidth]{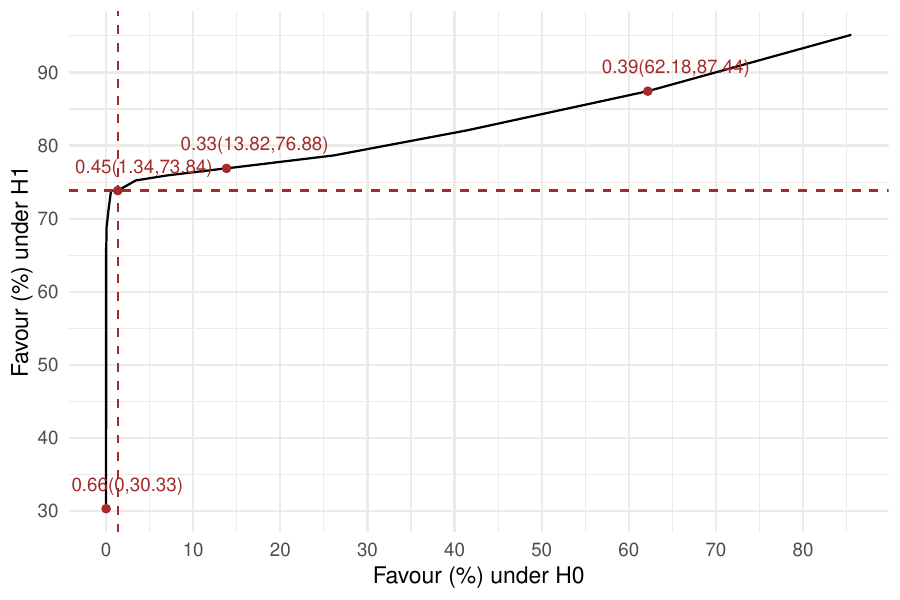}
         \caption{}
         \label{fig: threshold_s2_tradeoff}
     \end{subfigure}
          \begin{subfigure}[b]{0.5\textwidth}
         \centering
         \includegraphics[width=\linewidth]{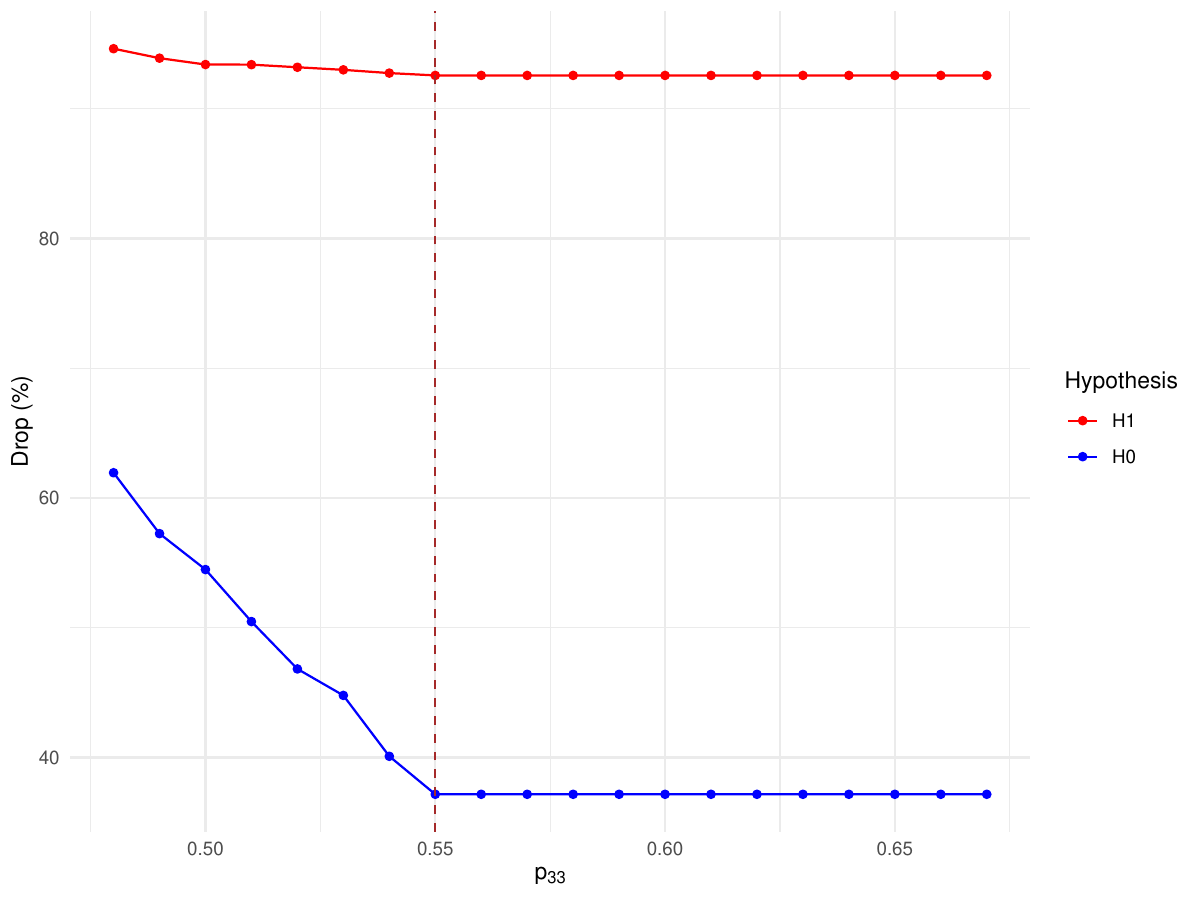}
         \caption{}
         \label{fig: threshold_s3_drop}
     \end{subfigure}

        \caption{Selecting probability thresholds for \textit{Mapped-$\alpha$} design. 
        (a) Trade off in favouring between the two hypotheses. The brown dashed lines intersect at the point of optimal trade off corresponding to threshold value 0.45 with the coordinates (x,y) representing the favouring under the null and alternate hypotheses respectively. 
        (b) Drop \% for different thresholds under the two hypotheses. The brown dashed line passes through the Drop \% values for the two hypotheses corresponding to the threshold 0.55.}
        \label{fig:threshold_selection}
\end{figure}

We express \textit{adaptability} as a percentage: at each stage, we replicate the \textit{Mapped-$\alpha$} design 10,000 times over a uniform grid of thresholds in $[0.33,0.66]$, and evaluate the proportion of simulated trials in which a specific adaptation (e.g., favouring $T_1$ or dropping $T_1$) is triggered under the two hypotheses. The grid is chosen considering that, for stage 2, the number of controls is fixed to 2 and there are 6 patients; therefore, the remaining probability of allocating the other two (active) arms is between $2/6=0.33$ and $1-(2/6)=0.66$. 
Figure~\ref{fig: threshold_s2_tradeoff} shows that the choice of $p_{21} = p'_{21}$ = $p''_{21}$ = 0.45 produces the best trade-off between the two competing hypotheses. This serves as the mid point of the decision line for stage 2.
For stage 3, we set the mid point of stage 3 decision line to the same value as decided for stage 2, i.e., $p_{32} = p_{32}' = p_{32}'' = 0.45$, and fix the dropping threshold $p_{31}$ to $\tau \in [0,0.2]$. The value of $\tau$ is the same value as taken in \textit{StratosPHere 2} and will be disclosed at the end of the trial to preserve study's integrity (see Section~\ref{sec: BRAR}). The threshold $p_{33}$ is chosen by simulation analyses similar to those performed for the previous stage: as shown in Figure~\ref{fig: threshold_s3_drop}, the probability of dropping the active arm under the Null is minimised for $p_{33} \geq 0.55$. 

For \textit{Mapped-$\beta$}, we should introduce the additional threshold by updating the values of $p_{2,1}'$ and $p_{3,2}'$ to $1/K$ where $K = 3$ (the number of arms); this means the region between $1/K$ and 0.45 represents the Balance region as defined before.

\paragraph{Operating Characteristics of the \emph{Mapped} vs. \emph{Unmapped} Designs}
A final part of addressing our research question (1) is to investigate: \textit{How do the operating characteristics of the original \textit{StratosPHere 2} design and other variants, including the \textit{Mapped} versions for our selected thresholds, compare against each other?} Results corresponding to the BRAR designs outlined in Table~\ref{tab:RAR_designs} are presented in Table~\ref{tab:mapping_OCs}. The operating characteristics are reported in terms of power, type-I error, and expected allocation probabilities under the alternative hypothesis of an optimal arm (see also Section~\ref{sec: Strato}). These are computed following a bootstrap approach (based on real data from a pilot \textit{StratosPHere 1} phase; see~\cite{jones2024stratosphere}), and by replicating the design a number of $10,000$ independent times. As shown in Table~\ref{tab:mapping_OCs}, the implemented {\it Mapping} rule not only guarantees a safe allocation (preventing undesirable allocation ratios), but also results in non negligible improvements both in terms of frequentist errors and participants allocated to the most promising arm.

\begin{table}[ht]
\centering
\begin{tabularx}{\textwidth}{ l|XX|XXX }
\toprule
  &\multicolumn{2}{c}{Frequentist properties} & \multicolumn{3}{|c}{Empirical Allocation}\\
  \midrule
Design & Power & Type-I error & Arm $C$ & Arm $T_1$ & Arm $T_2$ \\ 
  \midrule
  \textit{Fixed Equal Randomisation} & 0.748 & 0.12 & 0.34 (0.11)  & 0.33 (0.11)  & 0.33 (0.10)\\ 
\textit{Unrestricted} & 0.748 & 0.09 & 0.23 (0.12) & 0.23 (0.12) & 0.54 (0.17) \\ 
\textit{Control Protected} & 0.788 & 0.09 & 0.34 (0.07) & 0.20 (0.13) & 0.46 (0.13) \\ \hline
  \textit{StratosPHere 2} & 0.788 & 0.10 & 0.32 (0.07) & 0.21 (0.10) & 0.47 (0.12) \\ \hline 
  \textit{Permuted Block} & 0.783 & 0.12 & 0.30 (0) & 0.35 (0) & 0.35 (0) \\
  \textit{Mapped-$\alpha$} & 0.795 & 0.11 & 0.30 (0) & 0.20 (0.11) & 0.50 (0.11) \\ 
  \textit{Mapped-$\beta$} & 0.789 & 0.11 & 0.30 (0) & 0.20 (0.11) & 0.50 (0.11) \\ 
   \bottomrule
\end{tabularx}
\caption{Operating characteristics of the evaluated BRAR designs. Values are averaged across $10,000$ independent replicas; results are reported in terms of mean (standard deviation). Here, a value 0 for the standard deviation reflects the imposed restrictions on the number of arms. 
}
\label{tab:mapping_OCs}
\end{table}

\section{Handling Missing Data} \label{sec: MissingData}
In this section, we address research problem (2) as outlined in the Introduction. Specifically, we discuss here how to handle the occurrence of missing response data during the conduct of the \textit{StratosPHere 2} trial to effectively implement the (adaptive) \textit{Mapped-$\alpha$} design. Our main interest lies in the operating characteristics to assess how the BRAR is affected by potential missing data and different ways to handle it. To achieve this, we conduct simulation-based sensitivity analyses considering various cases of missing response data present only in the first two stages of the study. We do not include the possibility of missing response data in stage 3 of either of the strata, as this is the final stage of the trial and it does not guide any further adaptation. In particular, we want to understand whether and when an adaptation 
could still be implemented in the presence of missing data.

All simulations are carried out under a missing-at-random framework assuming  
all the components of the biomarker panel for a patient's primary endpoint are missing. We focus on the case where a maximum of 2 patients i.e. 10\% drop out of the study in the listed missing data cases. If the rate of missing data exceeds 10\%, we assume that no adaptations can be safely triggered. We also assume that when a data point is missing  we will refrain from imputing it as this is a small-sized trial. Imputation will be discussed later in this section. The following cases are considered and compared with a scenario with no missing data denoted as Case 0, i.e., when the response data is available for (6, 6, 8) patients in each stage respectively.
\begin{description}
    \item[Case 1] One patient's data is missing at random from stage 1 of a stratum, i.e., the composition of the stratum is (5, 6, 8) for the sample sizes of the first, second and third stages respectively. 
    \item[Case 2] Two patients' data are missing at random from stage 1 of a stratum, i.e., the composition of the stratum is (4, 6, 8) for the first, second and third stages respectively.
    \item[Case 3] One patient's data is missing at random from stage 2 of a stratum, i.e., the composition of the stratum is (6, 5, 8) for the first, second and third stages respectively.
    \item[Case 4] Two patients' data are missing at random from stage 2 of a stratum, i.e., the composition of the stratum is (6, 4, 8) for the first, second and third stages respectively.
    \item[Case 5] One patient's data is missing at random from each of stages 1 and 2 of a stratum, i.e., the composition of the stratum is (5, 5, 8) for the first, second and third stages respectively.
\end{description}

Results, in terms of the operating characteristics of the \textit{Mapped-$\alpha$} design evaluated under different cases of missingness, are reported in Table~\ref{tab:mapping_missing_OCs}, where the missing values were excluded from the simulations.
\begin{table}[ht]
\centering
\begin{tabularx}{\textwidth}{ lX|cc|ccc }
\toprule
  \multicolumn{2}{c|}{Missingness} &  \multicolumn{2}{c|}{Frequentist properties} & \multicolumn{3}{c}{Empirical allocation} \\
  \midrule
Scenario & \# Missing Data Points (per stage) & Power & Type-I error & Arm $C$ & Arm $T_1$ & Arm $T_2$ \\ 
  \midrule
Case 0 & (0, 0, 0) & 0.80 & 0.11 & 0.30 (0.0) & 0.20 (0.11) & 0.50 (0.11) \\   
Case 1 & (1, 0, 0) & 0.76 & 0.11 & 0.30 (0.2) & 0.20 (0.12) & 0.50 (0.12) \\
Case 2 & (2, 0, 0) & 0.75 & 0.10 & 0.30 (0.3) & 0.20 (0.13) & 0.50 (0.13) \\
Case 3 & (0, 1, 0) & 0.76 & 0.11 & 0.30 (0.2) & 0.20 (0.12) & 0.50 (0.12) \\
Case 4 & (0, 2, 0) & 0.74 & 0.11 & 0.30 (0.3) & 0.20 (0.12) & 0.50 (0.13) \\
Case 5 & (1, 1, 0) & 0.74 & 0.10 & 0.30 (0.4) & 0.20 (0.13) & 0.50 (0.13) \\
\bottomrule
\end{tabularx}
\caption{Operating characteristics of the \textit{Mapped-$\alpha$} design evaluated under different cases of missingness. The missing values were ignored in the simulations. Results are averaged across $10,000$ independent replicas and reported in terms of mean (standard deviation). Standard deviation of zero reflects the imposed restrictions on the number of control arms (always fixed to $2$).}
\label{tab:mapping_missing_OCs}
\end{table}
We note that with more missing data points, more power is reduced but overall, the operating characteristics remain reasonable allowing us to not interfere with the design. 

Now, we want to address the key question at the design stage -- whether to allow adaptation or not, when recommended by the BRAR design using the observed outcome data while ignoring any missing data. To investigate this, we utilise the parameter {\it adaptability} to observe the extent to which the design deviates from a balanced allocation in the simulations. 
Table~\ref{tab:missing_adaptability} illustrates how {\it adaptability} varies in the cases considered in this work. Since the data size is small, we wish to see lesser adaptability under the Null hypothesis so that under the Alternative, we can expect lesser adaptability and just let the trial design decide whether to adapt or not even during missingness without any interference. To decide whether to adapt or not during the interim analysis, we examine the situations one by one. During the first interim i.e., at the end of stage 1, we aim to understand how the adaptability varies for stage 2 when we have missing data points in stage 1 i.e., cases 1, 2 and 5 are possible. Referring to Table~\ref{tab:missing_adaptability}, we find that for these specific cases, adaptability under the the Null is very high (highlighted in bold in the table). Therefore, we are inclined to not adapt away from balanced allocation in the stage 2 if there is any missing datapoint in stage 1. Moving on to Interim 2, with a similar rationality, we have cases 3, 4 and 5 to consider and we note that the Drop/ Keep \% is very high under the Null. This leads us to not permit dropping either of the active treatment arms in stage 3 if missingness is found in stage 2. However, we may allow adaptation towards favouring an arm if need be. The final decisions on missing data handling in the ongoing trial are provided in the Statistical Analysis Plan of the trial \textit{StratosPHere 2}.  
\begin{table}[ht]
\centering
\begin{tabularx}{\textwidth}{ c|X|X|X|X|X|X }
\toprule
 & \multicolumn{2}{c|}{Stage 2 Favour/ Disfavour \%} & \multicolumn{2}{c|}{Stage 3 Favour/ Disfavour \%} & \multicolumn{2}{c}{Stage 3 Drop/ Keep \% } \\ 
\midrule
Scenario  & $H_0$    & $H_1$    & $H_0$    & $H_1$    & $H_0$    & $H_1$    \\ 
\midrule
Case 0   & 1.43  & 74.04 & 35.62 & 3.22  & 37.16 & 92.56 \\
Case 1 & \textbf{63.14} & \textbf{83.73} & 5.62  & 1.64  & \textbf{77.80} & \textbf{95.48} \\
Case 2 & \textbf{53.44} & \textbf{77.43} & 7.88  & 2.73  & \textbf{83.18} & \textbf{94.49} \\
Case 3 & 1.32  & 74.19 & 9.59  & 1.84  & \textbf{72.95} & \textbf{95.95} \\
Case 4 & 1.47  & 74.47 & 10.17 & 2.96  & \textbf{78.40} & \textbf{94.47} \\
Case 5 & \textbf{63.89} & \textbf{83.90} & 1.08  & 2.61  & \textbf{90.67} & \textbf{94.50}\\ 
\bottomrule
\end{tabularx}
\caption{Adaptability under the Null and Alternative hypothesis}
\label{tab:missing_adaptability}
\end{table}

\begin{figure}[ht]
    \centering
    \includegraphics[width=0.9\linewidth]{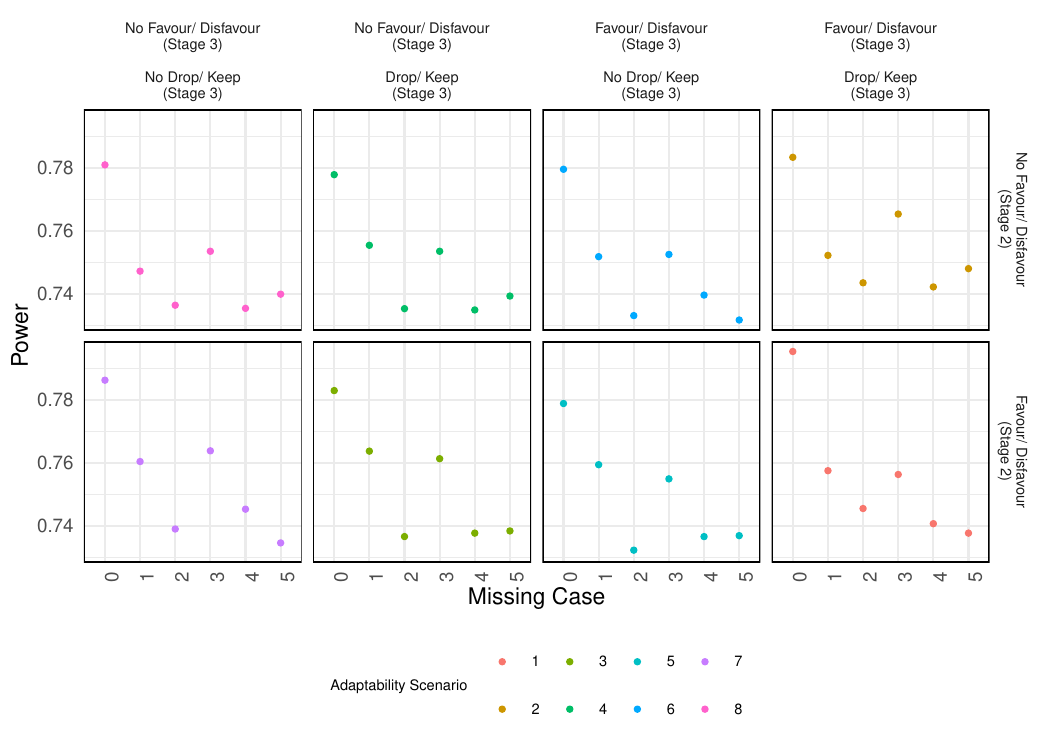}
    \caption{Power under different scenarios of adapting in the stage}
    \label{fig:missing_actions_power}
\end{figure}

Finally, we also review how the operating characteristics, particularly Power, would look like at the end of the study depending on our decision to adapt or not. This is illustrated in Figure~\ref{fig:missing_actions_power} representing Power under different scenarios of adaptability. We can see that, when data is complete (Case 0), adapting gives maximum power as expected. Note that some of these scenarios are redundant e.g., for missing case 3 where stage 1 has no missingness, it is pointless to consider the possibility of allowing favour in stage 2 as till that point data is complete and we do not interfere or interrupt. But, the plot helps us to understand how Power can get affected depending on what action we take. 

\paragraph{Imputation}
\cite{biswas2004missing} reports that based on simulations under the assumption of missing at random, statistical power improves when imputing missing data responses using the sample mean in their adaptive design as compared to not imputing. Here we assess how imputation affects the operating characteristics of the design. 

Given the specified trial design and the implementation of \textit{Mapped-$\alpha$}, we impute data only in the second stage of the trial with the sample mean of biomarker values observed up to the point of missingness for the corresponding missing arm in the following manner. Let \( \mathbf{y}_{i,k} \) be the vector of biomarker values for the \(i\)-th participant in the \(k\)-th arm, and \( \bar{\mathbf{y}}_k^{(t)} \) be the sample mean vector of the observed biomarker values in arm \(k\) up to time \(t\) (i.e., the timepoint at which the missingness occurs). If \( \mathbf{y}_{i,k} \) is missing in the second stage of the trial, it is imputed as 
$\mathbf{y}_{i,k} = \bar{\mathbf{y}}_k^{(t)} $ where $
\bar{\mathbf{y}}_k^{(t)} = \frac{1}{n_k^{(t)}} \sum_{m=1}^{n_k^{(t)}} \mathbf{y}_{m,k}$. Here, \( n_k^{(t)} \) is the number of observed biomarker vectors in arm \(k\) up to time \(t\). Other imputation methods along with experts' advice can be explored. Since this imputation can occur only when there is missingness in stage 2, only cases 3, 4 and 5 are relevant for comparisons. 

In Table~\ref{tab:missing_adaptability_imputation}, the bold text cells show the updated values reporting adaptability when missing data points are imputed only in stage 2. Other slight changes in the non-bold values are due to the variability of results due to simulations. We note that when missing data points are imputed, then under the Null hypothesis, the adaptability values for the cases 3 and 4 are similar to the complete case (Case 0) as expected as there is no missingness anymore. However, the same values for Case 5 are dissimilar to Case 0 since there still exists an unimputed missing point which is from stage 1. The updated values for cases 3 and 4 show the chances of dropping an active treatment arm under the null have decreased which is a favourable outcome suggesting that imputation can be recommended. These results show that when we want to be conservative with respect to the extent of adapting in case of missingness in the data, then imputing can also lower the scope of extreme adapting such as Drop/ Keep \% in stage 3. This observation is specific to our motivating trial but we believe it can be generalised to other studies highlighting the importance of checking levels of adaptability through simulations. However, the choice of imputation method will require a rigorous approach and expert advice.
\begin{table}[ht]
\centering
\begin{tabularx}{\textwidth}{ c|X|X|X|X|X|X }
\toprule
 & \multicolumn{2}{c|}{Stage 2 Favour/ Disfavour \%} & \multicolumn{2}{c|}{Stage 3 Favour/ Disfavour \%} & \multicolumn{2}{c}{Stage 3 Drop/ Keep \% } \\ 
\midrule
Scenario  & $H_0$    & $H_1$    & $H_0$    & $H_1$    & $H_0$    & $H_1$    \\ 
\midrule
Case 0  & 1.43  & 74.04 & 35.62 & 3.22  & 37.16 & 92.56 \\
Case 1 &  63.14 & 83.73 & 5.62  & 1.64  & 77.80 & 95.48 \\
Case 2 & 53.44 & 77.43 & 7.88  & 2.73  & 83.18 & 94.49 \\
Case 3 & 1.33  & 74.82 & \textbf{35.46}  & \textbf{3.26}  & \textbf{36.98} &\textbf{92.63} \\
Case 4 & 1.41  & 73.93 &\textbf{36.52 } & \textbf{3.49}  & \textbf{37.05 }&\textbf{92.60} \\
Case 5 & 63.61 & 84.31 &\textbf{5.49}  &\textbf{1.56}  & \textbf{78.27} &\textbf{95.76}\\ 
\bottomrule
\end{tabularx}
\caption{Adaptability under Imputation under the Null and Alternative hypothesis}
\label{tab:missing_adaptability_imputation}
\end{table}

\section{Further Analysis Challenges} 
\label{sec: FurtherChallenges}

\paragraph{Informative Missing Data}
In Section~\ref{sec: MissingData}, we addressed the issue of missing data from a design point of view. However, once the trial concludes, we will need to revisit the handling of missing data from an analysis perspective. First of all, the number of missing data cases to consider will increase as the missing data cases listed in Section \ref{sec: MissingData} do not account for missingness in stage 3 due to their lack of impact on adaptive decisions. 
Secondly, the pattern of missingness needs to be examined. In our simulations, we assumed that missing data occurred at random (MAR). This makes sense in our trial because of the nature of the endpoint. However, in other settings there is the possibility of missingness not at random (MNAR), where missing data may disproportionately occur in a specific treatment arm, signaling a non-random pattern. If this occurs, the risk of biased estimates increases, potentially compromising the trial’s operating characteristics. Therefore, identifying and understanding the missingness pattern is crucial. 
Finally, we will need to re-evaluate our imputation strategies and techniques, as those used previously may no longer be appropriate for the analysis stage of the trial. Before, in the design stage, we did not perform imputation in stage 1, and in stage 2 we imputed using historical data from the trial. However, after the conclusion of the study, if imputation is required, we can impute missing data in stage 1 also, as we now have more data points available to do so. A Bayesian imputation can be done to impute the missing data points.

\paragraph{Pooling Strata vs. Independent Analysis} In stratified designs or master protocols such as ``umbrella'', platform or basket designs, which are differentiated into multiple parallel sub-studies~\citep{park2019systematic}, investigators are often faced with the dilemma of pooling data from subgroups, borrowing some information, or simply following a set of parallel analyses. In fact, while conducting independent or \textit{stand-alone} analyses (such as the primary one in \textit{StratosPHere 2}) perfectly fit into the tailored paradigm of precision medicine, there are a number of concerns with this approach~\citep[see e.g.,][]{berry1990subgroup,berger2014bayesian}, including the issue of multiple testing and the lack of sufficient power. This is particularly relevant in rare disease trials, where the sample size is inherently low for detecting significant effects. One of the secondary analyses pre-planned in \textit{StratosPHere 2} is, in fact, a final analysis based on a pooled sample from the two mutation strata. Nonetheless, this may conflict with the potential heterogeneity of the treatment effects in the two strata, especially when the primary analysis (per strata) does not reach a decisive conclusion about the treatment effect(s) due to low power. Dedicated analyses should be conducted to assess when and to what extent a pooled analysis would be suitable and superior compared to independent stand-alone type of analyses: we report some preliminary exploration in Appendix~\ref{app: pooling}. Then, information borrowing principles, as done in e.g.,~\cite{zheng2022borrowing}, may be included to enhance the informative content of a stratum. In alternative, a decision rule could be identified to guide a \emph{``Pool vs. Don't Pool''} analysis, e.g., using a ``Test and Pool'' approach as discussed in~\cite{li2020revisit}. Such a decision rule may have several benefits in clinical research, starting from enhancing power and/or minimising the risk of incorrect conclusions (under strata and treatment heterogeneity) for the current trial, and ending with informing the design of future phases of the trial. In \textit{StratosPHere 2}, for example, it may suggest whether to keep a stratified approach in a future phase-2b or phase-3 trial.

\paragraph{Blinded vs. Veiled Analysis}
In placebo-controlled trials with multiple treatments that are noticeably different in terms of physical appearance or mode of administration, to ensure complete blinding (of both patients and physicians), a double-dummy approach must be implemented. That is, if we denote the two active treatments by A and B and their corresponding placebos by PA and PB, then each patient should receive two treatments having one of the following forms: {A, PB}, {B, PA}, {PA, PB}. In this way, neither patients nor physicians can be informed about the arm they are given or, most importantly, about the arm they are not receiving (as would occur in a standard placebo-controlled trial). This double-dummy technique is recommended by FDA \citep{FDA2023} for (confirmatory) platform trials but it can increase costs and place additional burdens on patients, potentially reducing compliance.
An alternative approach is based on having different placebos matching all the active treatments and administering a single arm from the following set: A, B, PA and PB. In this way, the patients remain unaware of whether they are receiving an active treatment or a placebo, but they do know which active treatment they have not been given. This partial blinding is termed Veiled in \cite{senn1995personal}.
The level of blinding ultimately leads to a re-definition of the hypothesis in Eq.~\eqref{eq: hypotheses}: i.e., should we compare the active treatment, say A, to the combined placebo control arm (i.e., both PA and PB), or only to the corresponding placebo arm PA? In \textit{StratosPHere 2}, we adopted a veiled approach and the control arm is represented by the combination of the two control arms (i.e., they are not differentiated). If we were to follow a double-dummy approach, we would need to differentiate between $C_{T_1}$ and $C_{T_2}$ and comparison should be made exclusively between $T_1$ vs. $C_{T_1}$ or $T_2$ vs. $C_{T_2}$. 
This also raises a fundamental question for the trial: how the veiled blinding compares with the double-dummy approach in terms of power as in the later the number of controls will increase? This can be answered by performing the simulations under two levels of blinding, however the ultimate blinding choice may not rely entirely on the statistical output and may consider the costs and other practical issues. 

\paragraph{Safety Reports}
In many trials, including adaptive trials, it is important to account for the differential nature of exposure to treatments. The patients will not have the same duration of taking treatments during the trial and these differences in follow-up times are not recorded in the commonly reported incidence proportion of an adverse event -- thus, introducing biases in estimating the adverse events by not accounting for time in treatment or follow-up. One way to address this is by updating the incidence proportions such that the duration of the treatment until an adverse event is factored in as suggested by \cite{allignol2016statistical}. More methods are provided by \cite{unkel2019estimands}. In this trial \textit{StratosPHere 2}, the drugs are repurposed, therefore, their safety information is already known to an extent. However, further work needs to be done while reporting safety analyses especially for adaptive trials where the follow-up times are bound to vary thereby, adding biases.

\section{Discussion} \label{sec: discussion}

In this article, we present research directly motivated by our involvement in conducting a stratified Bayesian Response-Adaptive Randomisation (BRAR) trial for a rare disease. We have explored methodological solutions to practical problems that limit the wider adoption of RAR in clinical trials. These specific challenges have been poorly addressed in the literature and continue to be a barrier to implementation. These specific challenges have been poorly addressed in the literature and continue to be a barrier to implementation in practice. In particular, we focus on two implementation issues that play a fundamental role in small-sample trials: (1) how can we minimise the chance of undesirable empirical treatment allocations while reflecting the theoretical allocation probabilities dictated by the RAR design and preserving an adequate level of the design's operating characteristics; and (2) how should we handle the occurrence of missing data in an RAR trial when decisions regarding adaptations need to be made at the interims?

In addressing (1), we have provided a general {\it Mapping} procedure that converts the vector of continuous allocation probabilities into a discrete allocation ratio. We have evaluated two instances of the proposed rule accounting for two levels of granularity for the final discrete set according to a different number of probability thresholds. We would like to emphasise that the evaluation was driven by the design characteristics of our motivating study, \textit{StratosPHere 2}. To decide the probability thresholds, we introduced an adaptability parameter recording deviations from the balanced allocation of the treatments. The thresholds have been appropriately selected in simulation studies by optimising the trade-off of the adaptability parameter under the Null and Alternative hypotheses. Note that the number of probability thresholds and their respective values can be adjusted with other practical considerations, e.g., based on statistical summaries of the allocation distribution or other specific constraints dictated by the unique requirements of the trial in hand.
The implemented {\it Mapping} rule also resulted in useful improvements in the operating characteristics of the trial, both in terms of frequentist errors and participants allocated to the most promising arms, when compared to the original \textit{StratosPHere 2} design. 

Additionally, we examined the impact of missing data on the operating characteristics of the trial and the newly introduced adaptability parameters. This analysis suggested the potential to adapt probability thresholds based on the extent of missingness, with a primary goal, for example, of minimising extreme adaptations such as the dropping of a treatment arm during stage 3 under the Null hypothesis (i.e., when all treatment are equivalent). However, for the purpose of the ongoing trial, we have opted to keep the probability thresholds as constant across the different scenarios of missing data. 
Further, the possibility of imputation is introduced and its impact on the trial design characteristics.

In conclusion, our solutions provide a way to conduct a safe trial (avoiding undesirable allocations and minimising wrong decisions in case of missing data) while serving as a strategy to enhance the frequentist properties of a small-sample trials and also keeping the essence of an RAR design. This procedure of {\it Mapping} is also amenable to a straightforward implementation for any type of randomisation system. To illustrate, Sealed Envelope \citep{SealedEnvelope2024} is a randomisation system commonly used by Clinical Trial Units that randomises patients to the treatment groups by utilising blocked randomisation list(s). In the \textit{StratosPHere 2} trial, the unblinded statistician is responsible for generating the final randomisation list(s) that includes permuted blocks containing treatments and allocation ratios that match the \textit{Mapped} design for BRAR. The unblinded statistician then provides the generated randomisation list(s) for the utilisation in the Sealed Envelope for patient assignments. Our experience with this trial shows that the \textit{Mapping} procedure can address the challenges of achieving the allocation ratio within small sample sizes while also simplifying the implementation for the randomisation system.
 Moreover, our findings suggest that relying solely on operating characteristics for handling missing data in an RAR design, especially at the design stage, could be insufficient. Incorporating additional parameters, such as the frequency of adaptations triggered during the trial, could provide deeper insights, ultimately leading to a more informed decision-making.

Finally, we highlighted four statistical problems for post-trial (final) analyses. We would have more missing data patterns, requiring an understanding of the missingness pattern and then the decision for the imputation method, if needed. Then, we raised the question of whether to pool data from different strata or not, which is especially relevant for rare disease early-phase trials. Next, we consider the impact of the level of blinding on the hypothesis testing. As a last point, we emphasise the importance of carefully thinking of how to best report adverse events by factoring in the varying follow-up times in adaptive trials. 


Directly motivated by our concrete experience in a stratified RAR trial for a rare disease, our proposals may warrant evaluation in broader contexts. This could include other trial designs or RAR rules. Further extensions could involve larger sample sizes or modifications to the number and size of trial stages. Another area for practical RAR exploration is missing data, particularly non-random missingness. While our analyses primarily focused on design and practical implementation, this work can be extended to the final analysis stage; crucially, the final analysis stage warrants exploration of additional imputation methods. The overall goal of this article was to describe practical challenges often neglected in technical literature and offer potential solutions for addressing them. We hope this inspires greater synergy between practical and methodological research, which is crucial for translating RAR's benefits into clinical practice.

\section*{Acknowledgement}
This research was supported by the UK Medical Research Council
$MC$\_$UU$\_$00002/15$ (SSV) and Efficient Study Design $MC\_UU\_00040/03$ (SSV) and Cambridge NIHR Biomedical Research Centre (MRT).

\section*{Conflict of Interests}
\noindent SSV is member of the advisory board for PhaseV. MRT received support from NIHR Cambridge BRC and MRC. He received consulting fees for advisory roles by Jansen, Apollo Therapeutics and Merck, and travel support from GSK and Jansen. He has been member of the ComCov and FluCov data safety monitoring/ advisory boards. This research is independent of these links. 

\bibliographystyle{apalike}

\appendix

\section{Appendix}

\clearpage

\subsection{Pooling Evaluation} \label{app: pooling}

To evaluate the potential of a pooled analysis, we consider the nine scenarios reported in Table~\ref{tab: pooling_scenarios}, and evaluate the corresponding operating characteristics (type-I error and power) of a \textit{stand-alone} analysis (i.e., independently by stratum, each based on $n=20$) vs. a pooled analysis (based on $N=40$). The evaluation is based on the \textit{StratosPHere 2} design, as summarised in Figure~\ref{fig: StratoPH}. 
\begin{table}[ht]
\centering 
\caption{Simulation scenarios with specification of the ``true'' treatment effects per stratum: $\mathbb{E}\Delta Y_{k} - \mathbb{E}\Delta Y_{C}$, for $k = T_1, T_2$. The bold font highlights the difference in the nonidentical scenarios.} \label{tab: pooling_scenarios}
  \begin{tabular}{lccccccc}
    \toprule
    \multirow{2}{*}{} &
      \multicolumn{3}{@{}@{}c@{}}{Stratum A} & &
      \multicolumn{3}{@{}@{}c@{}}{Stratum B} \\
      \cmidrule{2-4}\cmidrule{6-8}
      Scenario & $\mathbb{E}\Delta Y_{C}$  & $\mathbb{E}\Delta Y_{T_1}$ & $\mathbb{E}\Delta Y_{T_2}$ & & $\mathbb{E}\Delta Y_{C}$  & $\mathbb{E}\Delta Y_{T_1}$ & $\mathbb{E}\Delta Y_{T_2}$ \\
      \midrule
Scenario S1: identical strata (global null)  & 0.0 & 0.0 & 0.0 & & 0.0 & 0.0 & 0.0  \\  
Scenario S2: identical strata   & 0.0 & 0.0 & 0.3 & &0.0 & 0.0 & 0.3  \\ 
Scenario S3: identical strata   & 0.0 & 0.2 & 0.3 & &0.0 & 0.2 & 0.3  \\ 
Scenario S4: identical strata    & 0.0 & 0.3 & 0.4 & &0.0 & 0.3 & 0.4  \\ 
Scenario S5: similar strata    & 0.0 & 0.0 & \textbf{0.2} & &0.0 & 0.0 & \textbf{0.3}  \\ 
Scenario S6: similar strata    & 0.0 & 0.0 & \textbf{0.3} & &0.0 & 0.0 & \textbf{0.4}  \\
Scenario S7: divergent strata  & 0.0 & 0.0 & \textbf{0.0} & &0.0 & 0.0 & \textbf{0.3}  \\
Scenario S8: divergent strata   & 0.0 & \textbf{0.0} & \textbf{0.3} & &0.0 & \textbf{0.3} & \textbf{0.3}  \\
Scenario S9: divergent strata   & 0.0 & \textbf{0.3} & \textbf{0.0} & &0.0 & \textbf{0.0} & \textbf{0.3}  \\
    \bottomrule
  \end{tabular}
\end{table}

\begin{figure}[ht]
    \centering
    \includegraphics[scale = 0.58]{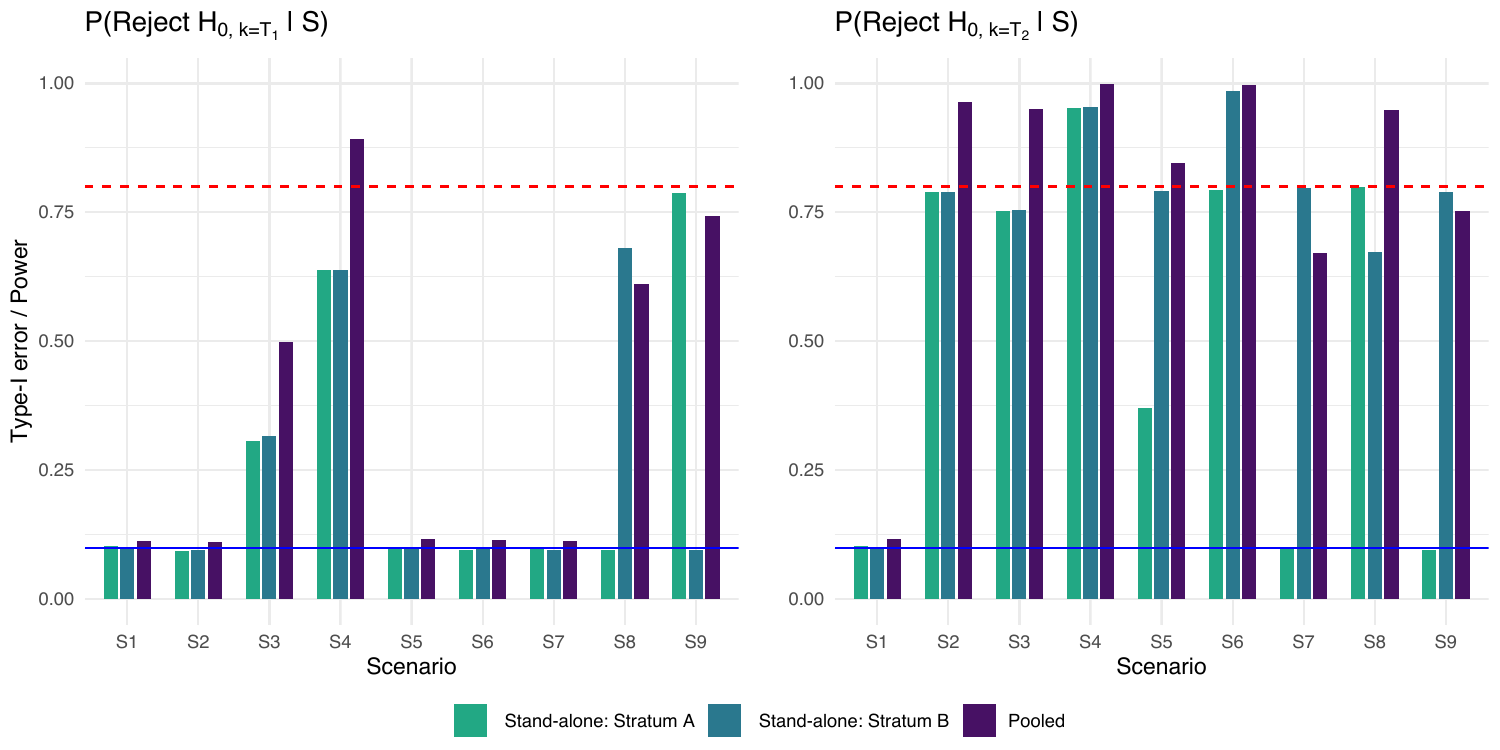}
    \caption{Type-I error and power (average results across 10,000 replicas) achieved by a stand-alone analysis by stratum ($n=20$) vs. a pooled analysis ($N=40$) in the evaluated scenarios. The horizontal lines indicate the targeted type-1 error ($\alpha = 0.1$; solid blue line) and power ($1-\beta = 0.8$; dashed red line).}
    \label{fig: type1-power-pooled}
\end{figure}

\begin{figure}[ht]
    \centering
    \includegraphics[scale = 0.6]{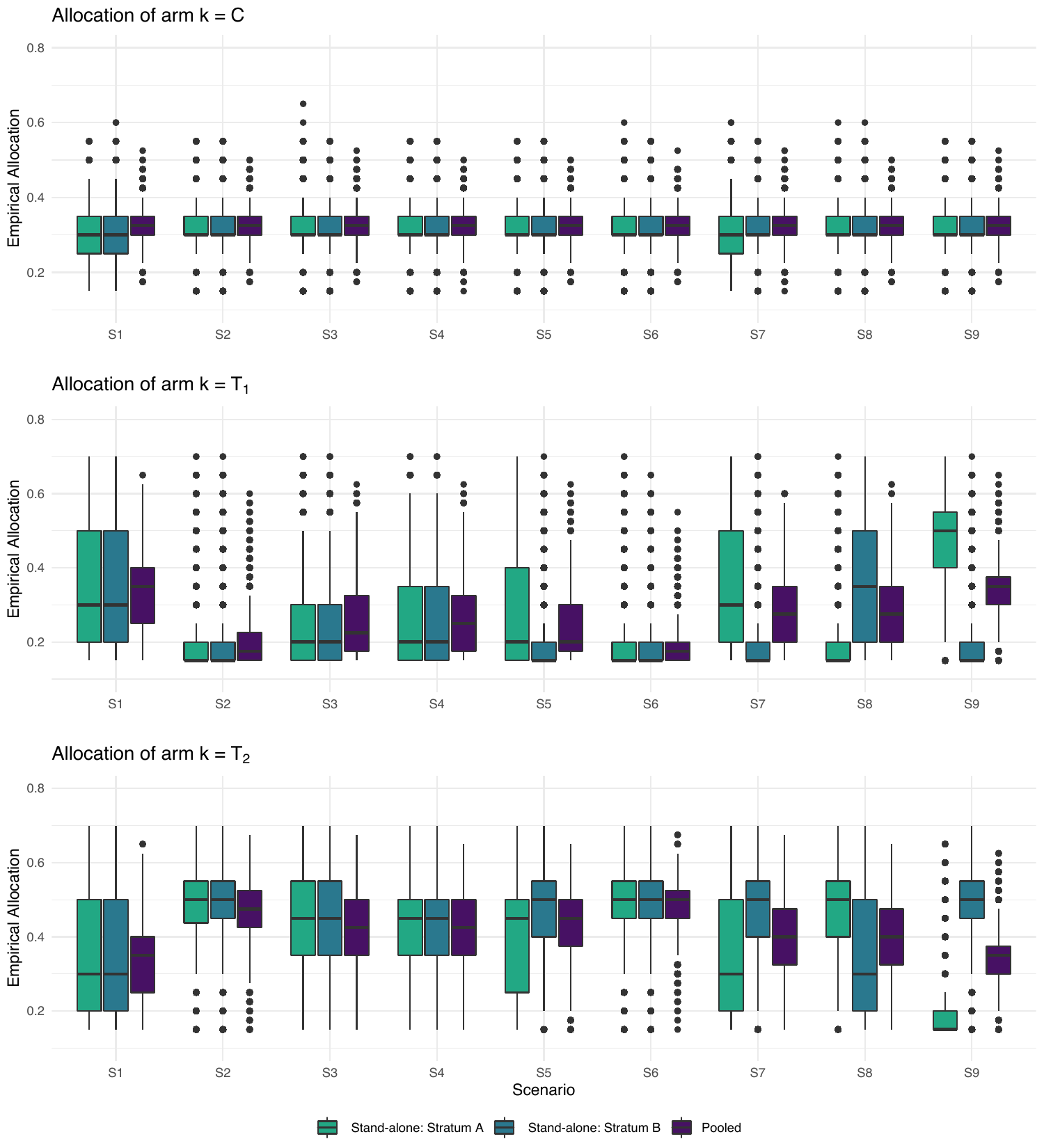}
    \caption{Proportion of the empirical arm allocation (box plots across 10,000 replicas) achieved by a stand-alone analysis by stratum ($n=20$) vs. a pooled analysis ($N=40$) in the evaluated scenarios.}
    \label{fig: alloc-pooled}
\end{figure}

As illustrated in Figure~\ref{fig: type1-power-pooled}, three different cases can be depicted. First, identical strata with at least one effective treatment (Scenarios S2-S4) greatly benefit from a pooled analysis, with substantial power improvements at a small type-I error inflation cost. The greatest advantages are encountered in scenarios with treatment effects that are too small to be detected with the trial sample (e.g., $T_1$ in Scenario S3), or when they are just small compared to the other evaluated treatments (e.g., $T_1$ vs. $T_2$ in Scenario S4). The latter is due to the underlying RAR design, which seeks to skew the allocation toward the most promising treatment, sacrificing the total allocation of the other active treatment (that is, $n_{T_1}$ in Scenario S3-S4; see Figure~\ref{fig: alloc-pooled}). Similar strata (Scenarios S5-S6) also show a power improvement, although to a smaller extent. Second, in divergent strata (Scenarios S7-S9), in addition to a potential negative impact on power, a pooled analysis can also affect interpretation: in S9, for example, a pooled analysis leads to a $75\%$ probability of rejecting the null of no treatment effect for both $k=T_1$ and $k=T_2$, whose true effect is 0 in exactly one of the two strata. The cases in which the two treatments show a divergent mechanism of action in the two strata clearly represent the worst-case scenarios for a pooled analysis. Finally, the global null case with identical strata and no treatment effects (Scenario S1) shows no major differences, except for negligible type-I error inflation.

\phantom{aaaa}

\end{document}